 \documentclass[final,3p,times,twocolumn]{elsarticle}

   \usepackage{graphicx}
\usepackage{amssymb}

\journal{Physics Letter B}

\begin{document}

\begin{frontmatter}



\title{Matrix elements of one-body and two-body operators between arbitrary HFB multi-quasiparticle states}


\author[1] {Qing-Li Hu}
\author[1]{Zao-Chun Gao\corref{cor1}}
\ead{zcgao@ciae.ac.cn}
\cortext[cor1]{Corresponding author}
\author[1] {Y. S. Chen}

\address[1] {China Institute of Atomic Energy, P.O. Box 275 (10), Beijing 102413, PR China}

\begin{abstract}
We present new formulae for the matrix elements of  one-body and
two-body physical operators, which are applicable to arbitrary
Hartree-Fock-Bogoliubov wave functions, including those for
multi-quasiparticle excitations. The testing
calculations show that our formulae may substantially reduce the
computational time by several orders of magnitude when applied to
many-body quantum system in a large Fock space.
\end{abstract}


\end{frontmatter}



\section{Introduction}
\label{s1}
Although the Schr\"{o}dinger equation was
proposed as early as in 1926, its exact solution (by means of the
full configuration interaction, FCI) for the quantum mechanical
many-body system is still hopeless except for the smallest system
due to the combinatorial computational cost. The mean-field theory
 has been a great success in describing the microscopic systems,
such as the nuclei, the atoms, and the molecules. The Hartree-Fock-Bogoliubov (HFB)
approximation, as the best mean-field method, has
played a central role in understanding interacting many-body quantum
systems in all fields of physics. However, the HFB wave functions
are far from the eigenstates of the Hamiltonian, and the effects
that go beyond mean-field are missing. Post-HFB treatments
(beyond-mean field methods), such as the configuration
interaction(CI), the generator coordinate method(GCM), and the
symmetry restoration, are expected to improve the wave functions and
present better description of the quantum mechanical many-body
systems. For instance, symmetry restoration of the HFB states has
been performed not only in the nuclei (e.g.\cite{Schmid04}), but
also in the molecules(e.g.\cite{Scuseria11}). Moreover, symmetry
restoration also improves the descriptions of quantum dots and
ultra-cold Bose systems in the condense matter world\cite{Yan07}.

The overlaps and the matrix elements of the Hamiltonian between the
HFB states are basic blocks to establish such post-HFB calculations.
Efficient evaluating of those quantities is of extreme importance to
implement the post-HFB calculations. Efforts have been devoted to
finding convenient formulae for such matrix elements and overlaps
for decades. The Onishi formula \cite{Onishi66,Ring80} is the first
expression of the overlap between two different HFB vacua, but the
sign of the overlap is not determined. Many works have been done to
overcome this sign problem
\cite{Hara79,Hara82,Neerg88,Ha92,Robledo94,Do98,Oi05,Be08,Robledo09}.
In Ref.\cite{Robledo09}, Robledo made the final solution and
proposed a new formula using the Pfaffian rather than the
determinant. After that, overlaps between quasi-particle states have
been intensively studied, which are also based on the Pfaffian
\cite{Robledo11,Oi12,Mizusaki12,Avez12,Bertsch12,Mizusaki13}. It is
realized that overlaps between multi-quasiparticle HFB states,
originally evaluated with the generalized Wick's
theorem(GWT)\cite{Balian69}, can be equivalently calculated by
compact formulae with
Pfaffian\cite{Oi12,Mizusaki12,Avez12,Bertsch12,Mizusaki13}. Thanks
to the same mathematical structure of the Pfaffian and the GWT, the
combinatorial explosion is avoided. We also should mention that,
before Robledo's work \cite{Robledo09}, there is another compact
formula for the GWT \cite{Perez07}. It is obtained by using Gaudin's
theorem in the finite-temperature formalism, but not expressed with
the Pfaffian.

Although the overlap between HFB states can be quickly calculated
using the proposed Pfaffian formulae or the method in \cite{Perez07}
to avoid the combinatorial explosion, one may certainly encounter
another difficulty in evaluating the matrix elements of many-body
operators, which has never been treated. We address this
problem as follows.
In the representation of second quantization, one can
write the one-body operator $\hat T$ and  two-body operator $\hat V$ as
\begin{eqnarray}\label{o1}
\hat T&=&\sum_{\mu\nu} T_{\mu\nu}\hat{c}^\dagger_\mu \hat{c}_\nu,\\
\hat
V&=&\frac14\sum_{\mu\nu\delta\gamma}V_{\mu\nu\gamma\delta}\hat{c}^\dagger_\mu
\hat{c}^\dagger_\nu \hat{c}_\delta \hat{c}_\gamma,\label{o2}
\end{eqnarray}
where $(\hat{c}^\dagger,\hat{c})$ are the creation and annihilation
operators of the spherical harmonic oscillator, i.e.
$\hat{c}^\dagger_\mu|-\rangle=|Nljm\rangle$ and
$\hat{c}_\mu|-\rangle=0$. $|-\rangle$ stands for the true vacuum.
Here, we assume all operators are defined in the same
$M-$dimensional Fock space.

The matrix element of an operator $\hat O$($=\hat T$ or $\hat V$)
 with multi-quasiparticle excitations is generally given as
\begin{eqnarray}\label{o}
\langle\Phi|\hat{\beta}_{i_1}\cdots\hat{\beta}_{i_L}\hat O \hat{ \mathbb{R}}
\hat{\beta}'^\dagger_{j_{L+1}}\cdots\hat{\beta}'^\dagger_{j_{2n}}|\Phi'\rangle,
\end{eqnarray}
where $\hat {\mathbb{R}}$
stands for a unitary transformation.
$|\Phi\rangle$ and $|\Phi'\rangle$ are different normalized
HFB vacua. $(\hat{\beta},\hat{\beta}^\dagger)$ and $(\hat{\beta}',\hat{\beta}'^\dagger)$ are
corresponding quasiparticle operators with
$\hat{\beta}_i|\Phi\rangle=\hat{\beta}'_i|\Phi'\rangle=0$ for any $i$.

Conventionally, the matrix element in Eq.(\ref{o}) can be obtained in two steps. The first step
 is evaluating the matrix element of each $c^\dagger_\mu c_\nu$
(or $\hat{c}^\dagger_\mu
\hat{c}^\dagger_\nu \hat{c}_\delta \hat{c}_\gamma$) in Eq.(\ref{o1}) [or  Eq.(\ref{o2})]
 through Pfaffian or the method in ref \cite{Perez07} to avoid the combinatorial explosion.
 The second step is collecting all the $c^\dagger_\mu c_\nu$
(or $\hat{c}^\dagger_\mu \hat{c}^\dagger_\nu \hat{c}_\delta
\hat{c}_\gamma$) matrix elements to get the final value of
Eq.(\ref{o}). Unlike the overlap between HFB states, each matrix
element of Eq.(\ref{o})(with $\hat O=\hat V$)
 requires the summation over $\mu,\nu,\delta,\gamma$.  This is too much time consuming for a symmetry restoration in a relatively
large configuration space, where thousands or millions of the matrix elements need to be calculated
 at each mesh point in the integral of the projection.
Such calculations in a large Fock space will be even too expensive
to be tractable.

In this Letter, we present new formulae for
evaluating the matrix elements of Eq. (\ref{o}) between arbitrary HFB states, which are in compact forms and
may greatly reduce the computational cost of the post-HFB calculations.

\section{Overlaps}
\label{theory}
Let's start with a useful equation that the expectation value of a product of
 arbitrary single-fermion operators, $\hat z_i$, is given by the Pfaffian of all possible contractions {\cite{Bertsch12,EL99,ER99}},
\begin{eqnarray}\label{m0}
\langle -|\hat{z}_1\cdots \hat{z}_{2k}|-\rangle=\mathrm{pf}(S),
\end{eqnarray}
where $S$ is a $2k\times 2k$ skew-symmetric matrix with the matrix
element $S_{ij}=\langle -|\hat{z}_i \hat{z}_j|-\rangle,\,
S_{ji}=-S_{ij},\,(i<j)$. One can extend Eq. (\ref{m0}) to a more
general form (details of proof are given in the Supplemental
material to this article),
\begin{eqnarray}\label{m}
\langle\Phi^a|\hat{z}_1\cdots
\hat{z}_{2n}|\Phi^b\rangle=\mathrm{pf}(\mathbb{S})\langle\Phi^a|\Phi^b\rangle,
\end{eqnarray}
where $|\Phi^a\rangle$ (or $|\Phi^b\rangle$) can be regarded as the
true vacuum or arbitrary
HFB vacuum. $\mathbb{S}$ is a $2n\times 2n$ skew-symmetric matrix,
but the matrix element in the upper triangular is
\begin{eqnarray}\label{sij}
\mathbb{S}_{ij}=\frac{\langle\Phi^a|\hat{z}_i
\hat{z}_j|\Phi^b\rangle}{\langle\Phi^a|\Phi^b\rangle}\quad(i<j).
\end{eqnarray}
For the lower triangular of $\mathbb{S}$,
$\mathbb{S}_{ji}=-\mathbb{S}_{ij}(i<j)$. Attention must be payed to
the useless contraction $\frac{\langle\Phi^a|\hat{z}_j
\hat{z}_i|\Phi^b\rangle}{\langle\Phi^a|\Phi^b\rangle}\,(i<j)$, which
never appears in the GWT and should not be taken as
$\mathbb{S}_{ji}(i<j)$. Here, we assume that
$\langle\Phi^a|\Phi^b\rangle$ is nonzero, and can be evaluated by
the available formulae proposed by several authors
 \cite{Robledo09,Oi12,Mizusaki12,Avez12,Bertsch12,Gao13}.

Here, we define the HFB vacuum $|\Phi^\sigma\rangle$ ($\sigma=a,b$)
as
\begin{eqnarray}\label{phi}
|\Phi^\sigma\rangle&=&\mathcal{N}_\sigma\hat\beta^\sigma_1\cdots\hat\beta^\sigma_{N_\sigma}|
- \rangle,
\end{eqnarray}
where $\mathcal{N}_\sigma$is the normalization factor of
$|\Phi^\sigma\rangle$. $N_\sigma$ is the number of $\hat
\beta^\sigma$ operators acting on $|-\rangle$ to form the HFB vacuum
$|\Phi^\sigma\rangle$. The operator $\hat z$ can be expressed in
terms of
 either $(\hat \beta^a,\hat \beta^{a\dagger})$ or $(\hat \beta^b,\hat \beta^{b\dagger})$,
\begin{eqnarray}\label{z}
\hat z_i=\sum_{j}\left(A_{ij}^a\hat\beta^a_j+B_{ij}^a\hat\beta^{a\dagger}_j\right)
=\sum_{j}\left(A_{ij}^b\hat\beta^b_j+B_{ij}^b\hat\beta^{b\dagger}_j\right).
\end{eqnarray}
We should stress that the coefficients  $A^a_{ij}$ and $B^a_{ij}$
(or $A^b_{ij}$ and $B^b_{ij}$) are arbitrary, which means $\hat z_i$
can stand for any single-fermion operator, such as $\hat c_i$, $\hat
c^\dagger_i$, $\hat \beta^a_i$, $\hat \beta^{a\dagger}_i$, $\hat
\beta^b_i$, $\hat \beta^{b\dagger}_i$, or even $\hat{
\mathbb{R}}\hat c_i\hat{ \mathbb{R}}^{-1}$, $\hat{ \mathbb{R}}\hat
\beta^{b\dagger}_i\hat{ \mathbb{R}}^{-1}$, etc. For instance, if
$\hat z_i=\hat \beta^a_i$, then $A_{ij}^a=\delta_{ij}$ and
$B_{ij}^a=0$.  The operators ($\hat c_i$, $\hat c^\dagger_i$),
($\hat \beta^a_i$, $\hat \beta^{a\dagger}_i$) and ($\hat \beta^b_i$,
$\hat \beta^{b\dagger}_i$) do obey the fermion-commutation
relations, but the general operator $\hat z_i$ does not have any
constraint. Hence, we do not impose $\hat z_i\hat z_j=-\hat z_j\hat
z_i$. By assuming the unitary  transformation between
$(\hat \beta^a, \hat \beta^{a\dagger})$ and $(\hat \beta^b, \hat
\beta^{b\dagger})$ being
\begin{eqnarray}\label{trans}
\left(\begin{array}{c}
\hat{\beta}^b\\\hat{\beta}^{b\dagger}
\end{array}\right)=\left(\begin{array}{cc}
\mathbb{X}&\mathbb{Y}\\\mathbb{Y}^*&\mathbb{X}^*
\end{array}\right)
\left(\begin{array}{c}
\hat{\beta}^a\\\hat{\beta}^{a\dagger}
\end{array}\right),
\end{eqnarray}
one can obtain the explicit expressions of $\mathbb{S}_{ij}$ in the
following three equivalent forms (see details in Supplemental
material),
\begin{eqnarray}\label{sija}
\mathbb{S}_{ij}&=&[A^aB^{aT}+A^a\mathbb{X}^{-1}\mathbb{Y}A^{aT}]_{ij},\\
\label{sijb}
\mathbb{S}_{ij}&=&[A^a\mathbb{X}^{-1}B^{bT}]_{ij},\\
\label{sijc}
\mathbb{S}_{ij}&=&[A^bB^{bT}+B^b\mathbb{Y}^*\mathbb{X}^{-1}B^{bT}]_{ij},
\end{eqnarray}
where the existence of the matrix $\mathbb{X}^{-1}$ is guaranteed
by the assumption $\langle\Phi^a|\Phi^b\rangle\neq 0$, according to
the Onishi formula \cite{Onishi66,Ring80}, in which
$\mathrm{det}\mathbb{X}\neq 0$ .

Note that Eq.(\ref{m}) can be regarded as a generalization of the
conclusion proposed recently in Ref.\cite{Mizusaki13}.

\section{Matrix elements of operators}
The matrix elements of Eq.(\ref{o}) can be rewritten in a general
form
\begin{eqnarray}
I&=&\langle\Phi^a|\hat{z}_1\cdots \hat{z}_{L} \hat O \hat{z}_{L+1}
\cdots\hat{z}_{2n}|\Phi^b\rangle\label{i},
\end{eqnarray}
where
\begin{eqnarray}
\hat{z}_k&=&\bigg\{\begin{array}{cc}{\hat{\beta}}_{i_k},&1\le k\le L\\
\hat{\mathbb{R}}{\hat{\beta}}'^\dagger_{j_k}\hat{\mathbb{R}}^{-1},&L+1\leq k\leq 2n \end{array} \\
|\Phi^a\rangle&=&|\Phi\rangle,\quad
|\Phi^b\rangle=\hat{\mathbb{R}}|\Phi'\rangle.
\end{eqnarray}
For fast calculation, we derive new formulae of $I$ instead of directly using Eq.(\ref{i}).
Here, we denote $I$ as $I_1$ for $\hat O =\hat T$, and $I_2$ for $\hat O =\hat V$.

To establish the notation, we define the following matrix elements
of $\mathbb{S}^{(\pm)}$ and $\mathbb{C}^{(\pm,0)}$,
\begin{eqnarray}\label{sc}
\mathbb{S}^{(+)}_{\mu k}&=&\bigg\{\begin{array}{cc}-\frac{\langle\Phi^a|\hat{z}_k \hat{c}^\dagger_\mu|\Phi^b\rangle}{\langle\Phi^a|\Phi^b\rangle},&1\le k\le L\\ \frac{\langle\Phi^a|\hat{c}^\dagger_\mu\hat{z}_k |\Phi^b\rangle}{\langle\Phi^a|\Phi^b\rangle},&L+1\leq k\leq 2n \end{array}, \\
\mathbb{S}^{(-)}_{\mu k}&=&\bigg\{\begin{array}{cc}-\frac{\langle\Phi^a|\hat{z}_k \hat{c}_\mu|\Phi^b\rangle}{\langle\Phi^a|\Phi^b\rangle},&1\le k\le L\\ \frac{\langle\Phi^a|\hat{c}_\mu\hat{z}_k |\Phi^b\rangle}{\langle\Phi^a|\Phi^b\rangle},&L+1\leq k\leq 2n \end{array},\\
\mathbb{C}^{(+)}_{\mu \nu}&=&\frac{\langle\Phi^a|\hat{c}^\dagger_\mu
\hat{c}^\dagger_\nu|\Phi^b\rangle}{\langle\Phi^a|\Phi^b\rangle},
\quad\mathbb{C}^{(-)}_{\mu \nu}=\frac{\langle\Phi^a|\hat{c}_\mu
\hat{c}_\nu|\Phi^b\rangle}{\langle\Phi^a|\Phi^b\rangle},\nonumber\\\\
\mathbb{C}^{(0)}_{\mu \nu}&=&\frac{\langle\Phi^a|\hat{c}^\dagger_\mu
\hat{c}_\nu|\Phi^b\rangle}{\langle\Phi^a|\Phi^b\rangle},
\end{eqnarray}
where the shapes of $\mathbb{S}^{(\pm)}$ and $\mathbb{C}^{(\pm,0)}$ are
$M\times 2n$ and $M \times M$, respectively.

For the one-body operator $\hat T$, we denote the quantity $T_0$ and the matrix $\mathbb{T}$ using above notations,
\begin{eqnarray}
T_0&=&
\sum_{\mu\nu}T_{\mu\nu}\mathbb{C}^{(0)}_{\mu\nu},\quad
\mathbb{T}_{ij}=\sum_{\mu\nu}T_{\mu\nu}\mathbb{S}^{(+)}_{\mu i
}\mathbb{S}^{(-)}_{\nu j}.
\end{eqnarray}
Similar to the Laplace expansion for determinant, there is also a
general expansion formula for Pfaffian (Lemma 4.2 in Ref
\cite{Stembridge90}, or Lemma 2.3 in Ref.\cite{Ishikawa99}). Due to
the same mathematical structure of the GWT and Pfaffian, this
Pfaffian expansion is essentially equivalent to the contraction role
of the GWT. We present several explicit expansions of Pfaffian in
the Supplemental material, and using the one with respect
to two rows (Eq.(S40) in Supplemental material) to get
\begin{eqnarray}\nonumber
&&\frac{I_1}{\langle\Phi^a|\Phi^b\rangle}=\frac{\langle\Phi^a|\hat{z}_1\cdots \hat{z}_{L} \hat T \hat{z}_{L+1}
\cdots\hat{z}_{2n}|\Phi^b\rangle}{\langle\Phi^a|\Phi^b\rangle}\\\label{ia2}
&=&T_0\mathrm{pf}(\mathbb{S})
-\sum_{i,j=1}^{2n}(-1)^{i+j+1} \alpha_{ij}\mathbb{T}_{ij} \mathrm{pf}(\mathbb{S}\{i,j\}),
\end{eqnarray}
where $\alpha_{ij}=1$ for $i<j$ and $-1$ for $i>j$. Here and below,
we denote $\mathbb{S}\{i,j,...\}$ as a sub-matrix of $\mathbb{S}$
obtained  by removing the rows and columns of $i$,$j$,$\cdots$.
  The indexes $i,j,\cdots$ are different from each other by definition.
  Thus we may set $\alpha_{ii}=0$, and hope this does not confuse the readers.

If $\mathrm{pf}(\mathbb{S})\neq0$, then $\mathbb{S}^{-1}$ exists.
pf$(\mathbb{S}\{i,j,...\})$ can be expressed with
 pf$(\mathbb{S})$ and some matrix elements of $\mathbb{S}^{-1}$ through the Pfaffian version of Lewis Carroll formula\cite{Ishikawa20}.
An alternative form of this formula has been given by Mizusaki and
Oi\cite{Mizusaki12} in the study of HFB matrix elements. Some
explicit expressions for this formula are given in the Supplemental
material. Here, we use the one for
$\mathrm{pf}(\mathbb{S}\{i,j\})$ (see Eq.(S54) in Supplemental material) to get
\begin{eqnarray}\label{ia4}
{I_1}=\left[T_0-\mathrm{Tr}(\mathbb{T}\mathbb{S}^{-1})\right] \mathrm{pf}(\mathbb{S}){\langle\Phi^a|\Phi^b\rangle},
\end{eqnarray}
where Tr is the trace of a matrix.

If $\mathbb{S}^{-1}$ does not exist, Eq.(\ref{ia4}) is  invalid, but one can compact Eq.(\ref{ia2}) to
\begin{eqnarray}\label{ia3}
I_1&=&\left\{T_0\mathrm{pf}(\mathbb{S})-\sum_{i=1}^{2n}\mathrm{pf}(\bar{\mathbb{S}}^i)\right\}{\langle\Phi^a|\Phi^b\rangle},
\end{eqnarray}
where the skew-symmetric matrices $\bar{\mathbb{S}}^i$ are the same as
$\mathbb{S}$ but the matrix elements in the $i$-th row and column $\bar{\mathbb{S}}^i_{ij}=-\bar{\mathbb{S}}^i_{ji}=\mathbb{T}_{ij}$.
[We set $\mathbb{T}_{ii}=0$ due to $i\neq j$ in Eq.(\ref{ia2})].

Calculation of the matrix element involving two-body operator is
more complicated.
Like the one-body operator $\hat T$, we define the following notations  associated with the
two-body operator $\hat V$,
\begin{eqnarray}
V_0&=&\frac{\langle\Phi^a|\hat
V|\Phi^b\rangle}{\langle\Phi^a|\Phi^b\rangle}=\frac14\sum_{\mu\nu\delta\gamma}V_{\mu\nu\gamma\delta}{\mathbb{C}_{\mu\nu\delta\gamma}},\label{v0}\\
\mathbb{V}^{(1)}_{ij}&=&\frac14\sum_{\mu\nu\delta\gamma}V_{\mu\nu\gamma\delta}{\mathbb{D}^{ij}_{\mu\nu\delta\gamma}},\label{v1}\\
\mathbb{V}^{(2)}_{ijkl}&=&\frac14\sum_{\mu\nu\delta\gamma}V_{\mu\nu\gamma\delta}{\mathbb{E}^{ijkl}_{\mu\nu\delta\gamma}},\label{v2}
\end{eqnarray}
where
\begin{eqnarray}
\mathbb{C}_{\mu\nu\delta\gamma}&=&\mathbb{C}^{(+)}_{\mu\nu}\mathbb{C}^{(-)}_{\delta\gamma}
-\mathbb{C}^{(0)}_{\mu\delta}\mathbb{C}^{(0)}_{\nu\gamma}
+\mathbb{C}^{(0)}_{\mu\gamma}\mathbb{C}^{(0)}_{\nu\delta},\\
\mathbb{D}^{ij}_{\mu\nu\delta\gamma} &=&\mathbb{C}^{(+)}_{\mu\nu}\mathbb{S}^{(-)}_{\delta i
}\mathbb{S}^{(-)}_{\gamma j}
-\mathbb{C}^{(0)}_{\mu\delta}\mathbb{S}^{(+)}_{\nu i}\mathbb{S}^{(-)}_{\gamma j}\nonumber\\
&+&\mathbb{C}^{(0)}_{\mu\gamma}\mathbb{S}^{(+)}_{\nu i}\mathbb{S}^{(-)}_{\delta j}
+\mathbb{C}^{(0)}_{\nu\delta}\mathbb{S}^{(+)}_{\mu i}\mathbb{S}^{(-)}_{\gamma j}\nonumber\\
&-&\mathbb{C}^{(0)}_{\nu\gamma}\mathbb{S}^{(+)}_{\mu i}\mathbb{S}^{(-)}_{\delta j}
+\mathbb{C}^{(-)}_{\delta\gamma}\mathbb{S}^{(+)}_{\mu i}\mathbb{S}^{(+)}_{\nu j},\\
\mathbb{E}^{ijkl}_{\mu\nu\delta\gamma}&=&\mathbb{S}^{(+)}_{\mu i}\mathbb{S}^{(+)}_{\nu j}\mathbb{S}^{(-)}_{\delta k}\mathbb{S}^{(-)}_{\gamma l}.
\end{eqnarray}
Similar to Eq.(\ref{ia2}), one can use Pfaffian expansions (Eq.(S40)
and Eq.(S52) in Supplemental material) to obtain the following $I_2$ expression,
\begin{eqnarray}\label{ib2}
&&\frac{I_2}{\langle\Phi^a|\Phi^b\rangle}=\frac{\langle\Phi^a|\hat{z}_1\cdots \hat{z}_{L} \hat V \hat{z}_{L+1}
\cdots\hat{z}_{2n}|\Phi^b\rangle}{\langle\Phi^a|\Phi^b\rangle}\nonumber\\
&=&V_0\mathrm{pf}(\mathbb{S})+\sum_{i,j=1}^{2n}(-1)^{i+j}\alpha_{ij}\mathbb{V}^{(1)}_{ij}\mathrm{pf}(\mathbb{S}\{i,j\})\nonumber\\
&+&\sum_{i,j,k,l=1}^{2n}(-1)^{i+j+k+l}\alpha_{ijkl}\mathbb{V}^{(2)}_{ijkl}\mathrm{pf}(\mathbb{S}\{i,j,k,l\}),
\end{eqnarray}
where $\alpha_{ijkl}=\alpha_{ij}\alpha_{ik}\alpha_{il}\alpha_{jk}\alpha_{jl}\alpha_{kl}$. Eq.(\ref{ib2}) clearly shows the contraction role of the GWT.

In analogy to Eq.(\ref{ia4}), if $\mathrm{pf}(\mathbb{S})\neq0$, by
replacing $\mathrm{pf}(\mathbb{S}\{i,j\})$ and
$\mathrm{pf}(\mathbb{S}\{i,j,k,l\})$ using the Pfaffian version of
Lewis Carroll formula (Eq.(S54) and Eq.(S55) in Supplemental material), one can
simplify Eq.(\ref{ib2}) as
\begin{eqnarray}\label{ib4}
&&{I_2}={\langle\Phi^a|\Phi^b\rangle}\mathrm{pf}(\mathbb{S})[V_0-\mathrm{Tr}(\mathbb{V}^{(1)}\mathbb{S}^{-1})\nonumber\\
&&+\sum_{i,j,k,l=1}^{2n}\mathbb{V}^{(2)}_{ijkl}
(\mathbb{S}^{-1}_{ij}\mathbb{S}^{-1}_{kl}-\mathbb{S}^{-1}_{ik}\mathbb{S}^{-1}_{jl}+\mathbb{S}^{-1}_{il}\mathbb{S}^{-1}_{jk})].
\end{eqnarray}
However, if $\mathrm{pf}(\mathbb{S})=0$, like Eq.(\ref{ia3}),
Eq.(\ref{ib2}) can be compacted to
\begin{eqnarray}\label{ib3}
I_2&=&\langle\Phi^a|\Phi^b\rangle\left\{V_0\mathrm{pf}(\mathbb{S})-\sum_{i=1}^{2n}\right.
\mathrm{pf}(\tilde{\mathbb{S}}^i)\\
&&\left.+\sum_{i,j=1}^{2n}(-1)^{i+j+1}\alpha_{ij}\sum_{k=1}^{2n}
\mathrm{pf}(\tilde{\mathbb{S}}^{ijk}\{i,j\})\right\},\nonumber
\end{eqnarray}
where $\tilde{\mathbb{S}}^i$ is the same as $\bar{\mathbb{S}}^i$
but $\mathbb{T}$ is replaced by  $\mathbb{V}^{(1)}$.
$\tilde{\mathbb{S}}^{ijk}$ is the same as $\mathbb{S}$ but the
matrix elements in the $k$-th row and $k$-th column
$\tilde{\mathbb{S}}^{ijk}_{kl}=-\tilde{\mathbb{S}}^{ijk}_{lk}=\mathbb{V}^{(2)}_{ijkl}$.

All the above formulae are based on the assumption
$\langle\Phi^a|\Phi^b\rangle\ne 0$. However, the case of
$\langle\Phi^a|\Phi^b\rangle= 0$ that leads to the well known Egido
pole \cite{Egido01} should be carefully studied. In this situation,
Eq.(\ref{m}) is invalid and Eq.(\ref{m0}) should be used. By
inserting Eq.(\ref{phi}) into Eq.(\ref{i}), and regarding  all
$\hat\beta^b$ and $\hat\beta^{a\dagger}$ as $\hat z$, one can
rewrite $I$ as
\begin{eqnarray}
I={\mathcal{N}_a\mathcal{N}_b}\langle - |\hat{z}_1\cdots
\hat{z}_{L'}
 \hat O \hat{z}_{L'+1} \cdots\hat{z}_{2n'}| - \rangle,
\end{eqnarray}
which is similar to Eq.(\ref{i}), but $L'=L+N_a$ and
$2n'=2n+N_a+N_b$. Although $I$ can be directly calculated with
Eq.(\ref{m0}) or the  formulae in Ref.\cite{Bertsch12}. However, one
can also derive corresponding compact forms in this situation.
Replacing $|\Phi^a\rangle$ and $|\Phi^b\rangle$ with $|-\rangle$, it
is seen all the above derived formulae from Eq.(\ref{sc}) to
Eq.(\ref{ib3}) are valid because $\langle-|-\rangle=1$. But, the
matrix $\mathbb{S}$ becomes $S$, whose shape is $(2n+N_a+N_b)\times
(2n+N_a+N_b)$, and much larger than the $(2n\times2n)$ dimension of
$\mathbb{S}$. Thus more computing time is required in this case.

\section{Discussions}
\label{calculation}

Numerical calculations have been performed to test the validity of
new formulae.
 The matrix elements of $\mathbb{S}$,
$\mathbb{S}^{(\pm)}$ and $\mathbb{C}^{(\pm,0)}$  are required and should
be evaluated with one of Eqs. (\ref{sija}-\ref{sijc}).
 Here, these matrix elements, together with $T_{\mu\nu}$ and
$V_{\mu\nu\gamma\delta}$, are chosen as complex random numbers. The
results show that the values of $I_1$ with Eqs. (\ref{ia4}), and
(\ref{ia3}) are indeed identical to that with the conventional
method. Similarly, the same values of $I_2$ with (\ref{ib4}),
(\ref{ib3}) and the conventional method are also confirmed 
(we present the testing FORTRAN code for $I_2$ in the Supplemental
 material).

\begin{figure}
  \includegraphics[width=3.5in]{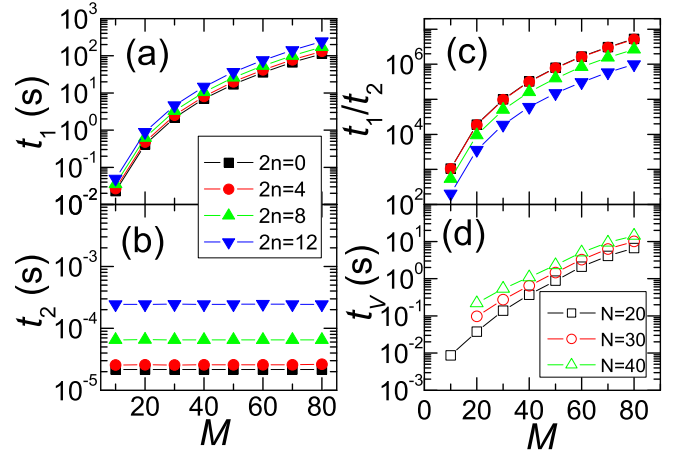}
  \caption{(color online) (a), CPU time, $t_1$, for the conventional method,
  as a function of $M$ and $2n$; (b), CPU time, $t_2$, for Eq.(\ref{ib4}),
  as a function of $M$ and $2n$, (c), Ratio of $t_1$ to $t_2$; (d)
  Total CPU time, $t_V$, for $V_0$, $\mathbb{V}^{(1)}$ and
  $\mathbb{V}^{(2)}$, $N$ is the dimension of $\mathbb{V}^{(1)}$ and
  $\mathbb{V}^{(2)}$ with $1\leq
i,j,k,l\leq N$.}   \label{time}
\end{figure}

The efficiency of the  most important Eq. (\ref{ib4}) is studied and
the results are shown in Fig.\ref{time}. Assuming $V_0$,
$\mathbb{V}^{(1)}$ and $\mathbb{V}^{(2)}$ are available, the
computational cost of Eq.(\ref{ib4}) is $O((2n)^4)$, which is
independent of $M$. This implies Eq.(\ref{ib4}) can be very
conveniently extended to large model spaces. In contrast, the
conventional method requires a time $O(M^4(2n)^3)$ which highly
depends on the model space due to the four-fold summation in
Eq.(\ref{o2}). Testing calculations have been carried out on a Intel
CPU with 2.4GHz. The elapsed time (in second), $t_1$ for the
conventional method and $t_2$ for Eq. (\ref{ib4}), are shown in
Fig.\ref{time}(a) and (b), respectively. To obtain the reliable
$t_1(t_2)$ value, identical calculations are repeated for many times
(denoted by $m$, ranging from 10 to $10^6$) until the total elapsed
time, $T$, is long enough, then $t_1(t_2)=T/m$. From
Fig.\ref{time}(c), the ratio $t_1/t_2$ can be easily above the order
of $10^6$ for $M=80$. Here, we chose $2n$ up to 12 because in the
practical calculations, it seems enough to include up to
6-quasiparticle states.

However, the elapsed time, $t_V$, for $V_0$, $\mathbb{V}^{(1)}$ and
$\mathbb{V}^{(2)}$ strongly depends on $M$. Moreover, $t_V$ is not
included in $t_2$ and should be separately considered. Fortunately,
all the $I_2$ matrix elements on top of the same ($\langle \Phi^a|$,
$|\Phi^b\rangle$) pair share the common $V_0$, $\mathbb{V}^{(1)}$
and $\mathbb{V}^{(2)}$. Thus they are evaluated just one time for
given HFB vacua, $|\Phi^a\rangle$ and $|\Phi^b\rangle$. Notice that
the computational cost of $\mathbb{V}^{(1)}$ and $\mathbb{V}^{(2)}$
also depends on their dimension, $N$, with $1\leq i,j,k,l\leq N$. To
cover all the $I_2$ matrix elements, $N$ should be properly chosen
in the range of $2n\leq N\leq2M$. Most of $t_V$ is taken by
$\mathbb{V}^{(2)}$, whose computational cost is $O(M^4N)$. The $t_V$
values for various $M,N$ are shown in Fig.\ref{time}(d). Comparing
with $t_1$, it looks that $t_V\approx 0.1t_1$ at large $M$. Let us
denote by $M_I$ the dimension of the $I_2$ matrix, and the global
efficiency of Eq.(\ref{ib4}) relative to the conventional method can
be evaluated through $r=\frac{M_I^2t_1}{t_V+M_I^2t_2}$. Suppose
$M_I=100,M=80$, $r$ can be easily in the order of $10^5$.

In Fig.\ref{time}(d), the CPU time, $t_V$ is within several seconds
for $M\leq80$, calculations may be implemented when one directly
uses Eq.(\ref{o2}), as is also taken in the standard $M-$scheme
shell model methods. However, $t_V$ can drastically increase with
$M$ bigger and bigger. Therefore, for heavy nuclei, one has to seek
a more concise form of two-body interaction, such as separable
interactions \cite{Tian09,Robledo10}, instead of directly using
Eq.(\ref{o2}). For instance, the Projected Shell Model (PSM) uses
the quadruple plus pairing interaction. The present method may be
conveniently applied to develop the PSM, so that it may includes the
states with more quasiparticles(e.g., 6-q.p., 8-q.p., etc).

\section{Summary}
\label{summary}

In this letter, we focused on the matrix elements of
one-body and two-body physical operators between arbitrary HFB
states. The formula of Eq.(\ref{m0}), used by Bertsch and Robledo
\cite{Bertsch12}, has been extended to evaluate the matrix element
of a product of single-fermion operators between two arbitrary HFB
vacua [see Eq.(\ref{m})]. Start from Eq.(\ref{m}), the matrix
elements of physical operators have been successfully transformed
into compact forms. Formulae for the pf$(\mathbb{S})=0$ case have
also been given. Besides, the case of the Egido pole with
$\langle\Phi^a|\Phi^b\rangle= 0$ has been discussed. Testing
calculations for the two-body operator matrix elements show that the
new formulae can easily be in several orders faster than the
conventional method. Thus those hopeless beyond mean field calculations
 for heavy nuclei in a large Fock space may be implemented by using the present method.

\textbf{Acknowledgements} Z.G. thanks Prof. Y. Sun and Dr. F.Q. Chen for the fruitful
discussions and the manuscript. The work is supported by the
National Natural Science Foundation of China under Contract Nos.
11175258, 11021504 and 11275068.

\appendix

\section{Supplemental material}

Supplementary material for mathematical details and the testing code can be found online at http://dx.doi.org/10.1016/j.physletb.2014.05.045.
\label{}







\end{document}